\documentclass[twocolumn,preprintnumbers,amsmath,amssymb]{revtex4}

\usepackage{graphicx}
\usepackage{dcolumn}
\usepackage{bm}

\begin{document}


\title{Combinatorial Quantum Gravity: Geometry from Random Bits}


\author{Carlo A. Trugenberger}
\email{ca.trugenberger@bluewin.ch}
\affiliation{%
SwissScientific, chemin Diodati 10, CH-1223 Cologny, Switzerland
}%

\date{\today}

\begin{abstract}
I propose a quantum gravity model in which geometric space emerges from random bits in a quantum phase transition driven by the combinatorial Ollivier-Ricci curvature and corresponding to the condensation of short cycles in random graphs. This quantum critical point defines quantum gravity non-perturbatively. In the ordered geometric phase at large distances the action reduces to the standard Einstein-Hilbert term. 
\end{abstract}
\maketitle

The Einstein equations of general relativity are the Euler-Lagrange equations of the Einstein-Hilbert action. In a quantum treatment, 
the Einstein-Hilbert action is perturbatively non-renormalizable. Within traditional quantum field theory the main approaches to resolve this conundrum are to postulate new physics at short scales and try to embed general relativity in a larger model, the string theory approach \cite{string}, or to look for a non-Gaussian ultraviolet (UV) fixed point that defines the theory non-perturbatively, the asymptotic safety approach \cite{safety}. This is perhaps best exemplified by the causal dynamical triangulations (CDT) program \cite{triang}, the gravity equivalent of lattice gauge theories, in which space-time is discretized in terms of (causal) simplicial complexes and the Einstein-Hilbert action formulated by Regge calculus \cite{regge}. 

In this paper I propose a different approach to quantum gravity and formulate a proof-of-concept toy model to show how this approach works. The idea is not to postulate space-time ab initio, but rather to consider it as an {\it emergent property} of purely combinatorial fundamental degrees of freedom. In this spirit, quantum gravity would be a close cousin of the Ising model: at short scales, physics is defined by an UV fixed point for fundamental constituents that are just random bits, the links of random graphs \cite{graphrev}; at large distances, the interaction is weaker and long-range order emerges in form of random geometric graphs \cite{geom}, which are random graphs equipped with a metric and define a discretization of a manifold. In particular, 
space and geometry are expected to emerge in the infrared (IR) limit due to the condensation of short graph cycles, the number of triangles being, e.g. one of the distinguishing features of random graphs vs. random geometric graphs \cite{krioukov}. As a driver of the quantum phase transition I will consider the combinatorial Ollivier-Ricci curvature \cite{olli1, olli2, olli3}, which becomes the standard Ricci curvature scalar in the ordered phase. Note that this program is totally different from previous approaches to quantum gravity based on graph structures \cite{foam, group}: there is no need  of auxiliary group variables and the action is a purely combinatorial version of the Einstein-Hilbert action. 

To show how this idea works concretely I will consider here a simplified toy model in which the configuration space CS is restricted to diluted random regular bipartite graphs. Bipartite graphs have no odd cycles, the smallest, ``elementary" loops being thus 4-cycles, squares. Two different squares on a graph can share zero, one or two edges (if they share three edges they must share also the fourth. i.e. they would be identical). By ``diluted" graphs I mean graphs in which two different elementary squares can share maximally one edge. This is a loop-equivalent  of a hard core requirement in a classical gas: the elementary constituent can touch but not overlap. Note that this restriction of the configuration space has nothing to do with any fundamental requirement, it is just a mathematical simplification that makes the toy model easily tractable. 

The partition function of the model is then defined by 

\begin{eqnarray}
Z &&= \sum_{\rm CS} {\rm exp} \left( -S_{EH}/\hbar \right) \ ,
\nonumber \\
S_{\rm EH} &&= - {1\over 2g} {\rm Tr}\  w^4  \ ,
\label{new1}
\end{eqnarray}
where $w$ denotes the adjacency matrices of the graphs in CS, $\hbar $ is the Planck constant and $g$ is the gravity coupling constant with dimension 1/action.  The dimensionless quantity $\hbar g$ will play the role of effective ``temperature" in this statistical field theory model. From now on I will focus on even connectivities of the regular graphs and denote these by $k=2d$.

Random regular bipartite graphs are "small worlds", i.e. their diameter and average distances on the graphs scale logarithmically with the number N of vertices (the volume) \cite{wormald}. They have locally a tree structure with very sparse short cycles governed by a Poisson distribution \cite{wormald} with mean $(2d-1)^l/l$ for cycles of length $l$. This behaviour is clearly not what is expected of a geometric space. As I now show, however, geometry emerges when the dimensionless coupling $\hbar g$ is small, the crucial point being that the interaction is nothing else than the {\bf discrete curvature scalar} for graphs. 

Random graphs are very different form simplicial complexes, which are regular configurations that can be always associated to a geometric realization. Because of their random character, the Regge formulation of curvature is no more applicable, a purely combinatorial version of Ricci curvature is needed. Recently, exactly such a combinatorial Ricci curvature has been proposed by Ollivier \cite{olli1} and further elaborated on in \cite{olli2, olli3}. 

As the continuum Ricci curvature is associated with a point and a direction on a manifold, its discrete version is associated with a vertex $i$ and a link $e_i$ of a graph. Averaging over all links emanating from a vertex gives the discrete version of the Ricci scalar at that vertex.  From a geodesic transport point of view, the Ricci curvature can be thought of as a measure of how much (infinitesimal) spheres (or balls) around a point contract (positive Ricci curvature) or expand (negative Ricci curvature) when they are transported along a geodesic with a given tangent vector at the point under consideration. The Ollivier curvature is a discrete version of the same measure. For two vertices $i$ and $j=i+e_i$ it compares the Wasserstein (or earth-mover) distance $W\left( \mu_i, \mu_j \right)$ between the two uniform probability measures $\mu_{i,j}$ on the spheres around $i$ and $j$ to the distance $d(i,j)$ on the graph and is defined as
\begin{equation}
\kappa (i,j)= 1- {W\left( \mu_i, \mu_j \right) \over d(i,j)} \ .
\label{olli}
\end{equation}
The Wasserstein distance between two probability measures $\mu_1$ and $\mu_2$ on the graph is defined as
\begin{equation}
W\left( \mu_1, \mu_2 \right) = {\rm inf} \sum_{i,j} \xi(i,j)d(i,j) \ ,
\label{wasser}
\end{equation}
where the infimum has to be taken over all couplings (or transference plans) $\xi(i,j)$ i.e. over all plans on how to transport a unit mass distributed according to $\mu_1$ around $i$ to the same mass distributed according to $\mu_2$ around $j$, 
\begin{equation}
\sum_j \xi (i,j) = \mu_1(i) \ , \qquad 
\sum_i \xi (i,j) = \mu_2(j) \ .
\label{transplan}
\end{equation}

The Ollivier curvature is very intuitive but, in general not easy to compute and work with. Fortunately, it becomes much simpler for bipartite graphs \cite{olli3}, which have no odd cycles. Since the Ollivier curvature of an edge depends only on the triangles, squares and pentagrams supported on that edge (a discrete form of locality) \cite{olli2} and there are no triangles and pentagrams on graphs in the configuration space, one can use for all practical purposes the simpler version of the Ollivier curvature for bipartite regular graphs \cite{olli3}:
\begin{eqnarray} 
\kappa (i,j) &&= -{1\over d} \Big[ (2d-2) -|N_1(j)| 
\nonumber \\
&&+ \sum_{a} \left( |L_a(j)|-|U_a(i)| \right) \times {\bf 1}_{\left\{ |U_a(i)| < |L_a(j)|\right\} } \Big] _+\ ,
\label{riccibipartite}
\end{eqnarray}
where $N_1(i)$ denotes the set of neighbours of $i$ which are on a 4-cycle supported on $(ij)$, ${\bf 1}$ denotes the indicator function (1 if the corresponding condition is satisfied, 0 otherwise) and the undescript ``+" denotes $z_+ = Max(z,0)$ so that the Ollivier Ricci curvature for bipartite graphs is always zero or negative. Suppose that $R(i,j)$ is the subgraph induced by $N_1(i) \cup N_1(j)$ and $R_1(i,j)$...$R_q(i,j)$are the connected components of $R(i,j)$. Then $U_a(i) = R_a(i,j) \cap N_1(i)$ and $L_a(j) = R_a(i,j) \cap N_1(j)$ for $a=1 \dots q$. 

This expression still looks forbidding but is, in reality quite simple. Two different squares (4-cycles) on a connected regular graph can either share 0 edges, if they are separated, or 1 edge or 2 edges if they touch. It is easy to convince oneself that the second term, involving the sum of connected components of a subgraph only contributes for squares that share 2 edges. Indeed, for an isolated square $|N_1|=1$ for all vertices on the square. 
If an edge supports $N_s$ squares which do not share another edge, then $|N_1(i)| = |N_1(j)|= N_s$ and $|U_a(i)|=|L_a(j)|$ since all the vertices within $N_1(i)$ and $N_1(j)$ are disconnected because of the absence of triangles in a bipartite graph and all the vertices of $N_1(i)$ are disconnected from those in $N_1(j)$ since, by assumption, the edge does not support two different squares. In the present model the Ollivier Ricci curvature reduces thus simply to 
\begin{equation}
\kappa (i,j) = -{1\over d} \big[ (2d-2) -N_s(ij) \big] \ ,
\label{riccireduced}
\end{equation}
where $N_s(ij)$ is the total number of squares supported on edge $(ij)$. Note also that I have left out the subscript ``+". This is because, for squares sharing maximally one edge $N_s(ij) \le (2d-2)$, as I now show. 

To do so, let me consider the uniform configuration with maximum square density.  First observe that, by the degree sum formula $2e = \sum_{i\ge3} i \ v_i$, with $e$ the number of edges and $v_i$ the number of vertices of degree $i$, one can derive that $2d$-regular graphs have exactly $dN$ edges. This means that one can uniquely assign to each vertex exactly $d$ edges. Out of $d$ edges one can form at most $d(d-1)/2$ different squares that share maximally one edge. Therefore the total number of squares is $Nd(d-1)/2$, each vertex having $d(d-1)/2$ squares uniquely assigned to it. Since a square is made of four vertices and four edges and there are a total of $N$ vertices and $dN$ edges, this means that each vertex is shared by exactly $2d(d-1)$ squares and each edge is shared by $2d-2$ squares. Thus, in this uniform configuration with maximum number of squares (sharing at most one edge) each edge supports exactly $2d-2$ squares, which shows that indeed $N_s(ij) \le 2d-2$. The maximum value $N_s(ij) = 2d-2$ for all edges is realized in Ricci flat, locally Euclidean graphs with neighbourhoods locally homeomorphic to  $\mathbb{Z}^d$.

The ``integral" of the Ollivier Ricci curvature scalar over the graph is 
\begin{eqnarray}
\sum_i \kappa (i) &&= -{2d-2\over d} N + {1\over d^2} \sum_i \sum_{e_i} N_s\left( e_i \right) 
\nonumber \\
&&= {-4\over d^2} \left[ {d(d-1)\over 2} N-N_s \right] \ ,
\label{totcursca}
\end{eqnarray}
with $N_s$ the total number of squares on the graph. The factor 4 comes from the fact that each square is shared by four vertices. 
On the other side, the total number of squares on a graph is given by \cite{squares} 
\begin{equation}
N_s = {1\over 8} \left[ {\rm Tr} \left( w^4 \right) - 8N d^2 + 2dN\right] \ .
\label{gensquares}
\end{equation}
Finally, one can combine (\ref{totcursca}), (\ref{gensquares}) and (\ref{new1}) to obtain
\begin{equation}
S_{EH} = -{d^2\over g} \Big[ \sum_i \kappa (i) + {6d-3\over d} N \Big] \ ,
\label{einstein}
\end{equation} 
which is a combinatorial version of the Einstein-Hilbert action (apart from an irrelevant constant). Indeed, sampling random regular bipartite graphs  according to the Boltzmann probability 
\begin{eqnarray}
p_B &&= {{\rm exp} \left( -S_{EH}/\hbar \right) \over \sum_{\rm CS} {\rm exp} \left( -S_{EH}/\hbar\right)} 
= {{\rm exp} \left( {d^2\over \hbar g} \sum_i \kappa (i) \right) \over Z} \ ,
\nonumber \\
Z &&= \sum_{\rm CS} {\rm exp} \left( {d^2\over \hbar g} \sum_i \kappa (i) \right) \ ,
\label{qugra}
\end{eqnarray}
amounts exactly to computing the combinatorial quantum gravity partition function. 

The free energy (divided by the statistical field theory ``temperature" $\hbar g$) is given by
\begin{equation}
F= {4\over \hbar g}  {d(d-1)\over 2} N   \big[ 1  -\zeta \big] -  S\left( N \right) \ ,
\label{freeenergy}
\end{equation}
where $\zeta = 2 N_s/(d(d-1)N)$ ($0\le \zeta \le 1$) is the density of squares and $S(N)$ the entropy of the graphs. 
The number $|{\cal G}^b_{N, 2d}|$ of random $2d$-regular bipartite graphs on $N$ vertices is known \cite{wormald},
\begin{equation}
|{\cal G}^b_{N, 2d}| = {\left(dN\right)! \ e^{-{1\over2}(2d-1)^2} \over \left( (2d)! \right)^N} \propto e^{dN {\rm ln} N} \ ,
\label{worm}
\end{equation}
for $N \gg d$.  
This would imply an entropy $S(N) = dN{\rm ln}N$ for large $N$. This number, however is drastically reduced by imposing the constraint of a finite density of squares $\zeta = O(1)$, as opposed to $\zeta = O(1/N)$ for completely random configurations. This is because the $d(d-1)N/2$ possible squares at each vertex are not independent degrees of freedom, since they have to combine to form a $2d$-regular bipartite graph. The same severe reduction in the number of allowed graphs is known to occur when passing from generic random graphs to random geometric graphs \cite{krioukov}. As I now show, the numerical evidence is that, at finite $\zeta = O(1)$  the entropy scales with a power of $N$ smaller than one. 

This has an important consequence. If $\hbar g$ is finite, the energy term will always dominate in this regime of finite square density and no phase transition can occur. Only if the ``temperature" $\hbar g$ is itself a growing function of $N$ can the 
energy-entropy balance tilt in favour of entropy at a critical point, where the fully random configurations in the CS start to dominate and random behaviour sets in. However, as I now show, this is exactly the relevant case for quantum gravity. 

To do so I will consider the continuum limit. In order to describe the emergent geometry in the continuum one has to assign a length $\ell $ to the links of the graph. I will take this length to scale as $\ell = \ell_0 N^{-1/d}$ so that the limit $N\to \infty$ represents the correct continuum limit. The constant $\ell_0$ can be interpreted as a fixed renormalization scale. On a graph that is locally homeomorphic to $\mathbb{Z}^d$, the combinatorial Ricci curvature between vertices $i$ and $j$ with geodesic tangent vector $v$ in the underlying Euclidean space $\mathbb{R}^d$ is given by \cite{olli1}
\begin{equation}
\kappa (i,j) = {\ell^2 R(v) \over 2(d+2)} + O\left( \ell^3 \right) \ ,
\label{cont}
\end{equation}
for $\ell \to 0$. Here $R(v)$ is the continuum Ricci curvature at $i=j$ in direction $v$. This result follows simply from the fact that, on a lattice and for $\ell \to 0$, the Wasserstein transportation measure becomes identical with the geodesic measure between balls used to define the continuum Ricci curvature. Integrating over direction and vertices one gets the limiting relation
\begin{equation}
{1\over \hbar g} \sum_i \kappa(i) \to {1\over 2(d+2) \ell_0^{d-2}} {N^{1-2/d}\over \hbar g} \int d^d x\ \sqrt{\eta} \ R \ ,
\label{ren}
\end{equation} 
where $\eta = 1$ is the determinant of the flat Euclidean metric, introduced here by hand only for completeness.

This expression is not entirely precise, as $N$ still appears in it: indeed, the 
crucial point here is to show that the model can be defined non-perturbatively and properly renormalized only if there is a second-order phase transition for the rescaled coupling $\hbar g/N^{1-2/d}$ rather than the original coupling $\hbar g$. This is because gravity is a model with a dimensionful coupling constant (for $d >2$) and thus $ \hbar g/N^{1-2/d}$ is the correct scaling of the true dimensionless coupling of the model, needed to compensate the fixed scale. 

This result, a ``temperature" $\hbar g$ growing with $N$, is exactly what we needed from an energy-entropy balance point of view to obtain a possible phase transition. In the strong gravity regime $\hbar g/N^{1-2/d} \gg1$, the energy (Euclidean action) term in the free energy is overwhelmed by the entropy and the typical configuration is that of a random regular bipartite graph. In this regime squares (and all other short cycles) are sparse, distributed according to a Poisson distribution with mean $(2d-1)^4/4$ , e.g. 600.25 for $d=4$ \cite{wormald} and graph distances scale logarithmically with the volume $N$. When gravity becomes weaker, $\hbar g/N^{1-2/d}\ll 1$, the energy term dominates the free energy and the typical configuration is one with the minimum energy, i.e. with the maximum number of squares $N_s =(d(d-1)/2)N$. This is a Ricci flat, locally Euclidean configuration with neighbourhoods homeomorphic to $\mathbb{Z}^d$ and graph distances scaling as $N^{1/d}$. In between these two extremal regimes one can expect a phase transition in which squares condense and geometric space emerges from a purely random configuration. 

The order parameter $N_s/6N$ corresponding to the relative number of squares is shown in Fig. 1 as a function of the rescaled parameter $\xi = \hbar g/\sqrt{N}$ for $d=4$ and $N=300$, $N=400$, $N=500$ in a Metropolis Monte-Carlo simulation. The correct scaling is evident and this result strongly suggests a second-order transition with critical point $\xi_c \simeq 500-1000$, which would define the model non-perturbatively.  The value $(3/4\pi) \xi_c \ell_0^2 /\hbar$ can then be identified with the gravitational constant $G_c$ at the critical point. Note that this result would also imply that the graph entropy in the geometric phase scales as
$S(N)\propto N^{2/d}$. Given that $N$ represents the (dimensionless) volume, this is a combinatorial version of the celebrated entropy area law \cite{area}. 

The model proposed here is only a ``solvable" toy model. A full treatment would require to derive the emergence of generic geometric graphs from generic random graphs when the combinatorial quantum gravity coupling becomes weak. Moreover, the issue of time and the Lorentzian signature of space-time has to be addressed. Concerning this point I would like to remark that, in this proposed approach to quantum gravity, there is nothing fundamental about time. There are only graphs at short distances (near the UV fixed point): space, time and also the difference between them (Lorentzian signature) are expected to emerge only at large scales. In a previous paper \cite{time} I have already explicitly derived a possible mechanism how a (3+1)-dimensional space-time with causal structure can form by quenching the fundamental graphs. The results presented here, although in a simplified version of the full model, lend strong support to the idea that the fundamental constituents of space-(time) are indeed random bits.

\begin{figure}
\includegraphics[width=8cm]{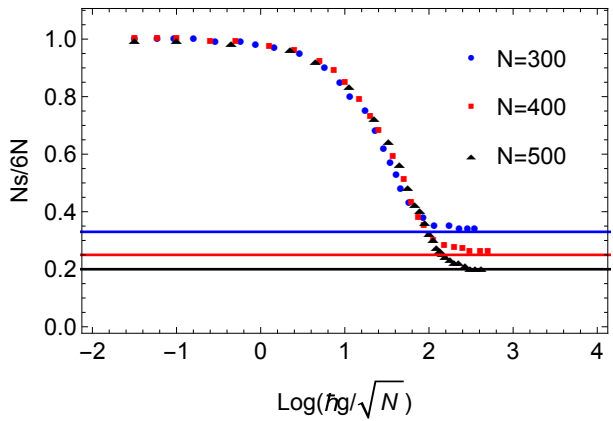}
\caption{\label{fig:Fig. 1} Monte Carlo simulation of the average number of squares for $d=4$ and $N=300$, $N=400$ and $N=500$ as a function of the rescaled coupling $\hbar g/\sqrt{N}$. Random regular graphs with sparse squares $N_s \sim {\rm Poisson\ } (600.25)$ and logarithmic distance scaling at large values of the coupling constant turn into $\mathbb{Z}^4$ lattices with the maximum number of squares $N_s=6N$ and power-law distances when gravitation becomes weak.}
\end{figure}


\end{document}